\documentclass[10pt]{article}

\usepackage{amsmath}
\usepackage{amssymb}
\usepackage{ulem}

\usepackage{graphicx}

\usepackage{cite}

\usepackage[labelfont=bf,labelsep=period,justification=raggedright]{caption}

\makeatletter
\renewcommand{\@biblabel}[1]{\quad#1.}
\makeatother

\date{}

\begin{document}

\begin{flushleft}
{\Large
\textbf{Coevolutionary immune system dynamics driving pathogen speciation}
}
\\
Kimberly J. Schlesinger$^{1,\ast}$, 
Sean P. Stromberg$^{1}$, 
Jean M. Carlson$^{1}$
\\
\bf{1} Department of Physics, University of California -- Santa Barbara, Santa Barbara, CA, USA
\\
$\ast$ E-mail: kschlesi@physics.ucsb.edu
\end{flushleft}

\section*{Abstract}
  We introduce and analyze a within-host dynamical model of the coevolution between rapidly mutating pathogens and the adaptive immune response. Pathogen mutation and a homeostatic constraint on lymphocytes both play a role in allowing the development of chronic infection, rather than quick pathogen clearance. The dynamics of these chronic infections display emergent structure, including branching patterns corresponding to asexual pathogen speciation, which is fundamentally driven by the coevolutionary interaction. Over time, continued branching creates an increasingly fragile immune system, and leads to the eventual catastrophic loss of immune control.
  

\section*{Introduction}
  
  The immune system is a complex adaptive system whose richness makes it an excellent model for nonlinear dynamics and biological complexity. From the basic physical and chemical interactions between foreign substances and the body's repertoire of lymphocytes, an array of complex system-wide behaviors can arise as the immune system works to recognize and eliminate harmful pathogens~\cite{owen_kuby_2013}. The development of immune system models has helped to identify mechanisms that underlie many of these emergent behaviors (e.g.~\cite{imm4phys,nowak_evolutionary_2006,sasaki94_drift,perelson2002modelling,wang06_hivPRL,chakraborty09_Tcells,stromberg12antia,ganusov2013mathematical}).
  
  The adaptive immune system of vertebrates has the remarkable ability to discriminate between self and non-self agents in the body, and to remove the foreign threats when recognized. The system consists of a complex array of lymphocytes, or white blood cells, which are able to recognize foreign agents with the high binding specificity of their receptors. These receptors are assembled randomly from gene segments in the bone marrow, and those that bind to the body's own cells are negatively selected as the lymphocytes mature in the thymus. The population of mature cells in the lymph nodes then has a diverse collection of specifically shaped receptors that can bind with high affinity to complementary peptide sequences, called epitopes, on many possible types of foreign antigen~\cite{owen_kuby_2013}. During an infection, lymphocytes that successfully bind with antigen rapidly proliferate to build an immune response that specifically targets the bound antigen for clearance.
  
  During the course of an infection, mutations that alter the shape, charge, or hydrophobicity of epitopes can impair continued recognition of the infection by the initially stimulated lymphocytes~\cite{erickson01_hepCepitope,allen04_escmutation}. Some rapidly mutating pathogens, most notably HIV, use this strategy to avoid clearance by the initial immune response and develop into a chronic infection~\cite{owen_kuby_2013,nowak_evolutionary_2006,johnson02_HIVevasion}. The adaptive immune system must then continuously adapt to control new mutant pathogen strains. This control can be aided by cross-reactivity: lymphocytes that bind strongly to one epitope can also bind with lower affinity to similarly shaped epitopes~\cite{welsh_heterologous_2010,stromb13_physbio}. Thus, a mutant with similar binding characteristics to the originally recognized epitope can be partially controlled by the existing immune response until a more specific response is stimulated~\cite{nowak_evolutionary_2006,kovsmrlj10_HLAspecificityHIV}. However, competition between lymphocytes, which during an infection swell to densities above the ideal homeostatic level, can also impair the overall immune response~\cite{wang06_hivPRL}. These dynamics of pathogen mutation and lymphocyte adaptation can be important in determining the eventual outcome of an infection.
 
  In this paper, we introduce a new model of this coevolution between the adaptive immune response and mutating pathogens. The model abstracts the chemical and molecular details of the binding interaction, while retaining important features that affect infection dynamics. We account for cross-reactivity by representing these populations on a phenotypic \textit{shape space}, in which the distance separating a pathogen and lymphocyte pair maps to a particular binding affinity (in general, larger distances give lower affinities). This method has been used in theoretical studies of immune system characteristics and behaviors affected by cross-reactivity, such as clonal repertoire size, self-nonself discrimination~\cite{ostperel79_ss}, and immunosenescence~\cite{stromb06_imsc}. We focus on the characteristics of T-cells, lymphocytes that form the primary response to certain rapidly mutating viruses such as HIV~\cite{primresp}.
      
\begin{figure}[ht!]
	\begin{center}
   \includegraphics[scale=1.0]{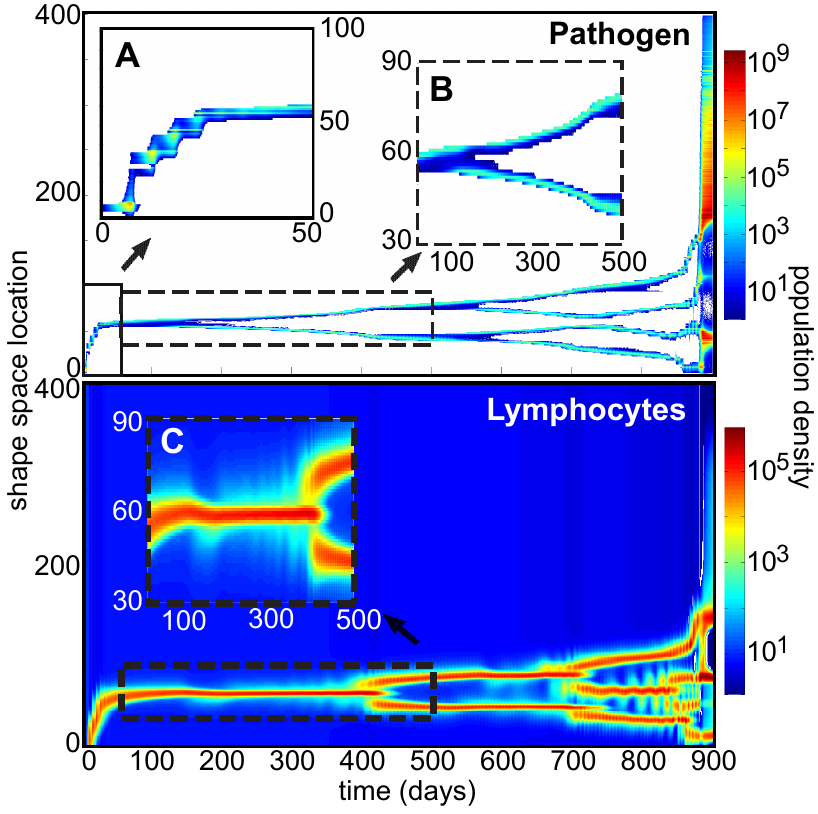}
	\end{center}
	\caption{ { \bf Coevolving shape space distributions of pathogen and T-cell populations over 900 days of infection.} The infection shown is an example of a stochastic mutating infection. Inset A (days 0-50) highlights the early periodic bursts of pathogen mutation which lead to chronic infection rather than clearance, driven by peaks in pathogen density which are subsequently controlled by the immune response. Insets B and C (days 50-500) provide a closer look at the chronic phase, which exhibits wave-like pathogen dynamics, pathogen speciation driven by the interaction with the immune response, and a resulting branch in T-cells, which is caused by selection alone. Following more branching, increased homeostatic pressure on the T-cells leads to the pathogens' dramatic escape from immune control (around day 870), marked by sudden rapid proliferation and unchecked diffusion throughout shape space. In this simulation, $\lambda=3.3~\mathrm{day}^{-1}$, $b=10~\mathrm{sites}$, $\Gamma=4~(\mu\mathrm{l}\cdot\mathrm{site}\cdot\mathrm{day})^{-1}$, $\xi = 2.3\times10^{-3}~\mathrm{day}^{-1}$, $\delta = 0.35~\mathrm{day}^{-1}$, and all other parameters are given in Table~\ref{params}.}
	\label{evol50}
\end{figure}

  In a rapidly mutating infection, the interactions we model give rise to complex dynamics. We observe several qualitatively different possible outcomes of infection, including early pathogen clearance, early pathogen escape, and the development of a long-lasting chronic infection. We also show the emergence of complex structure from the dynamics of a single chronic infection, including branching patterns in the pathogen population (Fig.~\ref{evol50}). In this evolutionary branching a unimodal phenotypic distribution of pathogens splits into two distinct and independently evolving clusters~\cite{doebeli00_branching}. This is analogous to asexual symptaric speciation~\cite{birkybarr09_asexspec} and arises naturally from the coevolutionary interaction between pathogens and T-cells. Although this speciation initially occurs while the infection is well-controlled, the numbers of lymphocytes needed to control the increasing number of distinct pathogen strains eventually exceeds the homeostatic constraints. This results in a sudden breakdown of immune control and a dramatic pathogenic escape, also visible in Fig.~\ref{evol50}.
  
\section*{Methods}

  To capture binding region diversity in the pathogen and T-cell populations, we describe them as densities denoted by $P(\vec{x},t)$ and $L(\vec{y},t)$ respectively~\cite{stromb06_imsc,stromb13_physbio}. The vectors $\vec{x}$ and $\vec{y}$ denote positions in a \textit{shape space} of phenotypes which determine the binding affinity $\gamma(\vec{x},\vec{y})$ between pathogens and T-cells~\cite{ostperel79_ss}. There is maximal binding complementarity when $\vec{x} = \vec{y}$, and monotonically decreasing affinity with increasing distance $|\vec{x}-\vec{y}|$ between T-cell and pathogen. Following previous theoretical work~\cite{segelper87,stromb06_imsc,stromb13_physbio}, we take this decay to be Gaussian:  
\begin{equation}\label{gamma}
\gamma(\vec{x},\vec{y}) = \exp\left[-\frac{|\vec{x}-\vec{y}|^2}{2b^2}\right].
\end{equation}
The parameter $b$ sets the specificity of antigen recognition and thus the length scale of the space. We do not consider the possibility of multiple epitopes, but identify each pathogen with a single shape space location. 

  The binding affinity mediates all interactions between pathogens and T-cells. The stimulation of T-cells by pathogens is modeled as a saturating function~\cite{stromberg_vaccination_2011} of pathogen density and proximity in shape space, a multiplicative factor ranging from zero to one:
\begin{equation}\label{Psat}
	P_{sat}(\vec{y},t) = \frac{\int \gamma(\vec{x},\vec{y}) P(\vec{x},t) d\vec{x}}{\kappa + \int 
		\gamma(\vec{x},\vec{y}) P(\vec{x},t) d\vec{x}} \in [0,1].
\end{equation}
There can be equivalent stimulation from low-density but high-affinity, high-density but low-affinity, or a combination of such pathogen distributions. If $P_{sat}(\vec{y},t)$ is high, T-cells at $\vec{y}$ are stimulated to divide and their decay is suppressed, generating an immune response.

  The killing of pathogens by T-cells is also a function of the affinity: the total killing rate of $P(\vec{x},t)$ is proportional to the \textit{effectivity},
\begin{equation}\label{omega}
\Omega(\vec{x},t) = \int\gamma(\vec{x},\vec{y})L(\vec{y},t)d\vec{y},
\end{equation}
a measure of the quality of an immune response~\cite{stromb13_physbio}.

  The following coupled differential equations describe pathogen and T-cell dynamics:

\begin{eqnarray}
\frac{dL(\vec{y},t)}{dt} &=& \Gamma + \xi \left[1-\frac{L_{tot}(t)}{R}\right] L(\vec{y},t) + \sigma P_{sat}(\vec{y},t)L(\vec{y},t) - \delta \left[1-P_{sat}(\vec{y},t)\right] L(\vec{y},t); \label{dL-basic} \\
\frac{dP(\vec{x},t)}{dt} &=& \left[1-\frac{P_{tot}(t)}{\phi}\right]\int \lambda(\vec{x}')P(\vec{x}',t)Q(\vec{x}',\vec{x})d\vec{x}' - \beta P(\vec{x},t) \Omega(\vec{x},t).\label{dP-mut}
\end{eqnarray}

  The first term of Eq.~(\ref{dL-basic}) represents a constant influx $\Gamma$ of na\"{i}ve T-cells from the bone marrow or thymus. The second term, describing logistic growth, accounts for homeostatic competition between all T-cells~\cite{sprentsurh,stromberg_suppression_2010}, independent of binding characteristics. This allows proliferation under lymphopenic conditions, and decay when the total T-cell density $L_{tot}(t) = \int L(\vec{y},t)~d\vec{y}$ is above the carrying capacity $R$, as is usually the case during response to an infection. The third term of Eq.~(\ref{dL-basic}) describes T-cell proliferation in response to stimulation by pathogens; this occurs at rate $\sigma \times P_{sat}(\vec{y},t)$, which falls between 0 and $\sigma$ depending on the pathogen population surrounding $\vec{y}$. T-cells turn over at rate $\delta$ (fourth term of Eq.~(\ref{dL-basic})), but in activated T-cells this turnover is suppressed by the factor of ($1-P_{sat}$). 

  In the absence of immune response (i.e.~$\Omega = 0$), Eq.~(\ref{dP-mut}) describes pathogens proliferating at rate $\lambda(\vec{x})$. We account for a fitness landscape for the pathogens (independent of immune pressure) by making this rate a function of $\vec{x}$. Since all pathogens compete for the same resources, we include a logistic factor in the first term of Eq.~(\ref{dP-mut}), limiting growth of any individual strain as the total pathogen density $P_{tot}(t) = \int P(\vec{x},t)\, d\vec{x}$ approaches a capacity $\phi$. It has been shown that target cell models for viral dynamics reduce to a logistic equation~\cite{stromb12_exhst}.
  
     \begin{table}[!ht]
\caption{Parameters used in all simulations unless otherwise noted. Values are approximated within biologically relevant ranges, based on known immune system characteristics and previous modeling work~\cite{stromb13_physbio,stromberg12antia} (see Appendix S1 for details). Although the exact phase diagram boundaries between regions with different infection outcomes may change at different parameter values, the qualitative dynamics within each infection outcome are not especially sensitive to the exact values of these parameters.}
\begin{tabular}{| l | c | c |}
	\hline
	Parameter & Sym. & Value\\
	\hline
	pathogen mutation rate & $\mu$ & $2.2\times10^{-5}~(\mathrm{base}\cdot\mathrm{cycle})^{-1}$ \\
	binding specificity & $b$ & $20~\mathrm{sites}$ \\
	pathogen growth & $\lambda$ & $3~\mathrm{day}^{-1}$ \\
	pathogen capacity & $\phi$ & $10^{10}~\mu\mathrm{l}^{-1}$ \\
	pathogen killing & $\beta$  & $10^{-5}~\mu\mathrm{l}\cdot\mathrm{day}^{-1}$ \\
	na\"{i}ve cell influx & $\Gamma$ & $1~(\mu\mathrm{l}\cdot\mathrm{site}\cdot\mathrm{day})^{-1}$ \\
	homeostatic pressure & $\xi$ & $6.1\times10^{-4}~\mathrm{day}^{-1}$ \\
	T-cell replication & $\sigma$ & $3~\mathrm{day}^{-1}$ \\
	T-cell decay & $\delta$ & $0.33~\mathrm{day}^{-1}$ \\
	stimulation coefficient & $\kappa$ & $10^{5}~\mu\mathrm{l}^{-1}$ \\
	\hline
\end{tabular}
\label{params}
\end{table}
  
  We incorporate mutation into the logistic growth law of Eq.~(\ref{dP-mut}) with a mutation kernel $Q$~\cite{nowak_evolutionary_2006}. The matrix element $Q(\vec{x},\vec{x}')$ gives the mutation rate of a pathogen from shape space location $\vec{x}$ to location $\vec{x}'$, as a fraction of the replication rate $\lambda(\vec{x})$. This mutation conserves pathogen number:
\begin{equation}\label{conserve}
	\int Q(\vec{x},\vec{x}')d\vec{x}' = 1,~~~\forall~~\vec{x}.
\end{equation} 

  In this study, we choose a decaying kernel, so that the mutation rate between two different sites falls off as distance between the sites increases:
\begin{equation}
Q(\vec{x},\vec{x}') = \left(\frac{2}{\pi}\right)^{\frac{1}{2}}\chi |\vec{x}-\vec{x}'|^{-2}, ~~~ \forall~~ \vec{x}\neq\vec{x}'. 
\end{equation}  
Mutations to nearby sites are thus much more likely than long-distance ones, but long-distance mutations still occur at a non-negligible rate, due to the possibility that a single amino acid substitution may substantially change the charge or hydrophobicity of an entire binding region. Finally, for our simulation lattice of $n$ sites, we choose $\chi$ so that the total fraction of pathogen at $\vec{x}$ that mutates to any \textit{different} site $\vec{x}' \neq \vec{x}$ sums to the overall pathogen mutation rate $\mu$:
\begin{equation}\label{mu}
	\sum_{\vec{x}'\neq\vec{x}} Q(\vec{x},\vec{x}') = \mu,~~~\forall~~\vec{x}.
\end{equation}

  In order to understand the roles of particular parameters in the outcome of these infections, we represent the process as deterministic. We can also extend this to a stochastic model by shuffling the kernel, generating new set of rates $Q(\vec{x},\vec{x}')$ at regular intervals of 0.1 days. To do this, we draw a rate for each $\vec{x} \neq \vec{x}'$ from a folded normal distribution with standard deviation $\chi |\vec{x}-\vec{x}'|^{-2}$, and set the rates for $\vec{x} = \vec{x}'$ to satisfy Eq.~(\ref{conserve}). The series of kernels generated has a time average equal to the original kernel. Although in this study we focus mainly on the deterministic model, we include some results of this stochastic generalization, which are explicitly noted when presented.
  
  Our simulations are performed on a one-dimensional lattice with $n = 400$ sites. We hold pathogen fitness $\lambda(\vec{x})$ constant across shape space, except at the edges of the lattice where it converges to zero to prevent edge effects. The population density at any site is automatically set to zero upon falling below a threshold of 1 to prevent unrealistic dynamics resulting from arbitrarily low densities. Parameter values are provided in Table~\ref{params}; further information about the estimation of these parameters is provided in Appendix S1.
  
\section*{Results}

Before it is inoculated with pathogens, the immune system is at rest at homeostatic equilibrium. The corresponding steady-state solution with no pathogens in the system is a uniform distribution of T-cells across shape space, which models well the random distribution of na\"{i}ve cells before exposure to infection:

	\begin{equation}
	P^*(\vec{x}) = 0, ~~~\forall ~~\vec{x}; ~~~ L^*(\vec{y}) = \frac{R}{2n}\left[\left(1-\frac{\delta}{\xi}\right)+\sqrt{\left(1-\frac{\delta}{\xi}\right)^2+\frac{4n\Gamma}{R\xi}}\right], ~~~ \forall ~~\vec{y}.
	\end{equation}		

	We choose the carrying capacity $R$ to equal the total equilibrium population of T-cells ($R = n \times L^*(\vec{y})$). This results in the simpler steady-state solution which we use for our initial condition:
	
	\begin{equation}\label{cleared_sstate}
	P^*(\vec{x}) = 0, ~~~\forall ~~\vec{x}; ~~~ L^*(\vec{y}) = \frac{\Gamma}{\delta}, ~~~ \forall ~~\vec{y}.
	\end{equation}		

	This pathogen solution is not stable, but we assume that a scenario with no pathogens in the system exists before inoculation. During an infection, the presence in our model of a threshold population density below which all pathogen populations are assumed to be 0 allows total clearance to occur. The T-cell solution is a stable attractor: as long as no pathogens exist in the system, the T-cells will converge to carrying capacity $R = \Gamma n / \delta$ with naive cells uniformly distributed through shape space. 

  During an infection, however, the T-cell population must be elevated above carrying capacity for some amount of time to establish control over the pathogens. The cost of this growth of the T-cell population above homeostatic equilibrium is a slow decay of T-cells across all of shape space, at a rate proportional to the amount by which the population exceeds the equilibrium. This reduces both the growth rate of stimulated T-cells and the diversity of the overall T-cell population, and can severely compromise the immune system's ability to control an infection.

  At time $t=0$, we inoculate our system with a small, localized dose of pathogens at site $\vec{x}_0$. Since the initial dose grows exponentially until large enough to either (a) stimulate the immune response, (b) produce viable mutant strains, and/or (c) near its own carrying capacity, the exact size of the dose does not qualitatively affect the trajectory of the infection, as long as it is small enough to avoid triggering these processes initially. We use the value $P(\vec{x}_0,t=0) = 10~\mu \mathrm{l}^{-1}$.
  
\begin{figure*}[!ht]
	\begin{center}
	\includegraphics[scale=0.92]{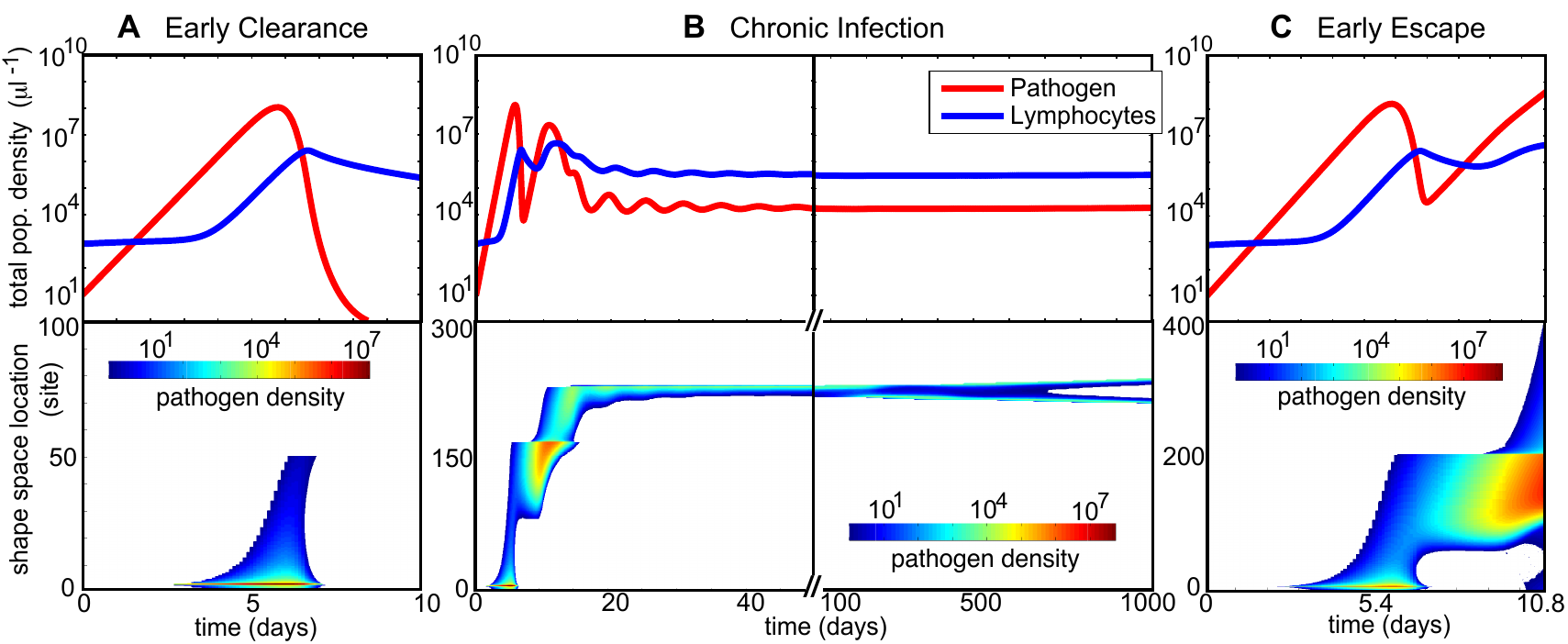}
	\end{center}
	\caption{ {\bf Examples of three of the four distinct types of infection observed in the deterministic model. } (The case of sterilizing immunity 
is not shown.) The total population densities of both pathogens and T-cells (top panels), as well as the shape space distribution of the pathogens (bottom panels), are plotted over time for each infection. A.~Early clearance after the acute phase of infection ($\mu = 1\times 10^{-5}~(\mathrm{base}\cdot\mathrm{cycle})^{-1}, b = 36 \mathrm{sites}$). B.~Chronic infection in which mutant pathogen strains avoid clearance by the initial immune response ($\mu = 2.5\times 10^{-5}~(\mathrm{base}\cdot\mathrm{cycle})^{-1}, b = 32~\mathrm{sites}$). The first 50 days, in which this initial avoidance occurs, are shown on the left; on the right (days 50 - 500), the total pathogen density remains nearly constant while slow antigenic drift and branching occur. C.~Early escape ($\mu = 3\times 10^{-5}~(\mathrm{base}\cdot\mathrm{cycle})^{-1}, b = 24~\mathrm{sites}$). Some pathogens are cleared by the initial immune response, but the population reaches carrying capacity without being controlled. Parameter values in Table~\ref{params} were used unless otherwise noted.}
	\label{4types}	
\end{figure*}

  Over a range of biologically reasonable parameter values, we find that there are four possible qualitatively distinct trajectories that a single infection may take (depicted in Fig.~\ref{4types}):

\begin{itemize}

\item[(1)] In regions of parameter space where $\Omega(\vec{x}_0,t=0) > \lambda / \beta$, the pathogen population decays immediately upon introduction to the system, an outcome known as \textit{sterilizing immunity}. In this paper, we focus on infections well outside of these regions, and this outcome is not shown in Fig.~\ref{4types}.

\item[(2)] Also possible is \textit{early clearance}, in which the pathogen population initially grows exponentially but is completely cleared before reaching a steady-state level (Fig.~\ref{4types}A). 

\item[(3)] In some regions of parameter space, an infection is able to escape early clearance, often by mutating to establish itself in shape space locations outside the reach of the initial immune response. These infections then approach an approximate steady state, where the pathogens coexist with the immune response at a controlled level in a localized region of shape space for an extended amount of time, which we term a \textit{chronic infection} (see Fig.~\ref{4types}B). A chronic infection may also include mutations, leading to the antigenic drift of the localized pathogen population. 

\item[(4)] Finally, the pathogens may avoid not only early clearance, but also any lasting localization or reduction of their population by the immune response (see Fig.~\ref{4types}C). These \textit{early escape} infections grow to carrying capacity or fill the entire shape space before the immune system can establish control.

\end{itemize}

\subsection*{Development of Chronic Infection: Role of Pathogen Mutation and Immune Trade-offs}

\begin{figure*}[!ht]
 	\begin{center}
    \includegraphics[scale=0.92]{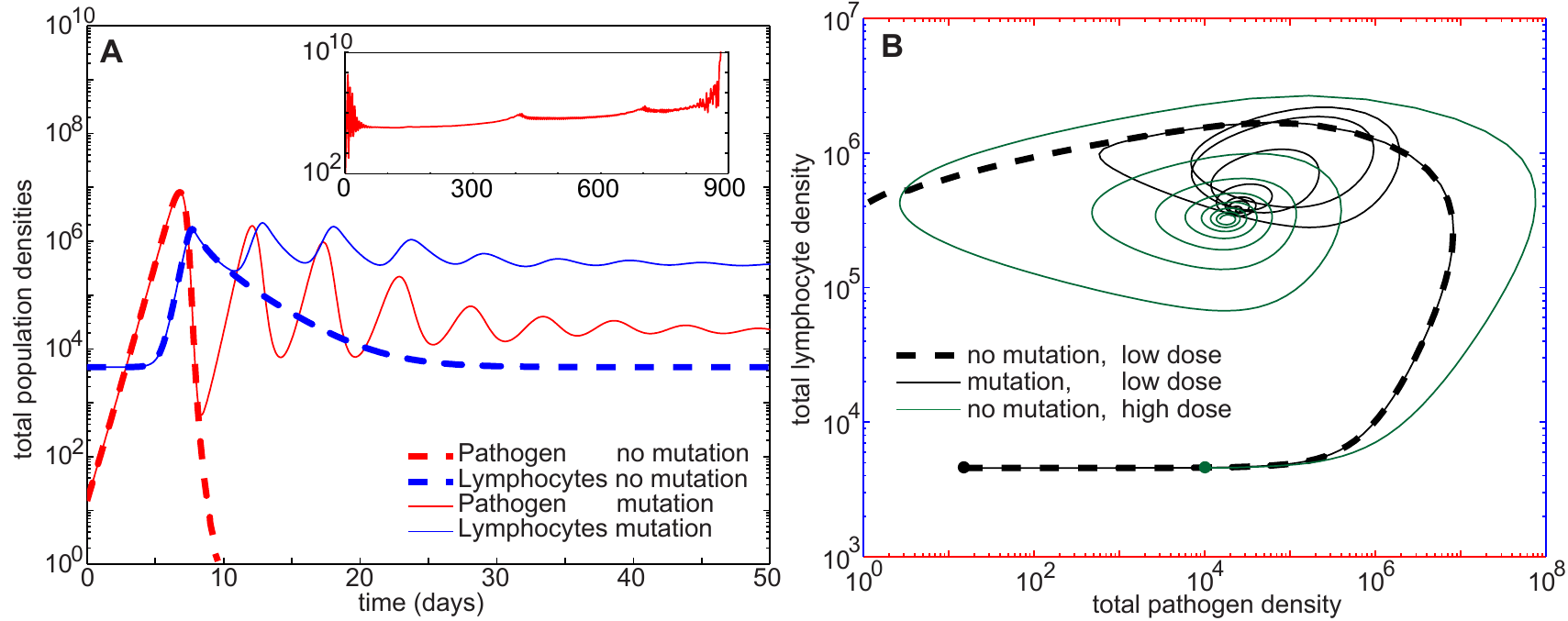}
    \end{center}
 	\caption{ { \bf Coevolutionary dynamics in the first 50 days of infection. } Shown here are the mutating infection from Fig.~\ref{evol50} and a non-mutating infection with the same parameter values. A.~Plots of total pathogen and total T-cell populations for both non-mutating (dashed) and mutating infections (solid curves). When there is no mutation, the pathogen density decays below the threshold and is set to zero. With mutation, pathogens escape clearance as strains arise with lower affinity to the T-cell response (see Fig.~\ref{evol50}A). Both pathogens and T-cells then converge to an approximate equilibrium state corresponding to a chronic infection. The inset shows total mutating pathogen density for the entire 900-day infection, which displays qualitative features observed in HIV and SIV~\cite{mudd12_HIVvaccine}; peaks in the chronic phase correspond to branching events (see Fig.~\ref{evol50}). B.~Phase plot of total T-cell versus total pathogen densities for the first 50 days of the two simulated infections in A and a third infection with a higher initial dose of pathogens. Of the low-dose infections, only the mutating case escapes clearance, converging to a chronic state similar to the approximate steady state (Eq.~\ref{chronic_sstate}) existing in the non-mutating system. Parameters are given in Table~\ref{params}.}
  	\label{four-leaf}
\end{figure*}

  With these initial conditions, the system undergoes dynamics corresponding to the acute phase of infection: roughly exponential growth of pathogens until the T-cells are stimulated, followed by the growth of T-cells until the pathogen population begins to decay. Examples for both mutating and non-mutating infections during this phase (days 0-50) are shown in Fig.~\ref{four-leaf}. Any system trajectory which eventually brings the pathogen population at all sites below the threshold -- for example, the black dotted curve in Fig.~\ref{four-leaf}B -- will lead to total pathogen clearance, and a subsequent return of the system to the equilibrium state in Eq.~(\ref{cleared_sstate}). This occurs in the non-mutating infection in Fig.~\ref{four-leaf}A, where pathogens are cleared about a week after inoculation. 
  
  The mutating pathogens in Fig.~\ref{four-leaf}A, however, with the same initial conditions and the same system parameters except for mutation rate, are able to avoid early clearance by generating new strains with low affinity to the initial immune response. (These strains are visible in Figs.~\ref{4types}B and~\ref{evol50}A.) This allows the T-cell and pathogen densities to converge to a chronic infection, an approximate equilibrium at which T-cells coexist with a controlled, localized pathogen population. 

	The corresponding steady-state solution can be approximated from our equations in the case of non-mutating pathogens localized at a single site $\vec{x}_0$. With the assumption that the T-cell population is large enough early in the infection that the influx $\Gamma$ is negligible, the pathogen population's steady-state value in the non-mutating system is
	\begin{equation}\label{abits}
	P^*(\vec{x}_0) \approx \frac{\kappa(\delta - \xi(1-L_{tot}/R))}{\sigma + \xi(1-L_{tot}/R)}.
	\end{equation}
	The terms beginning with $\xi$ in the numerator and denominator of Eq.~\ref{abits} are negligible in estimating the steady state as long as $\sigma R/\xi >> |L_{tot}-R|$ and $\delta R/\xi >> |L_{tot}-R|$, which is a good approximation for all infections in this study. In the simplest case, where we assume the T-cells are also localized at a single shape space site $\vec{y}_0$ in order to approximate the narrow peaks they form under the homeostatic constraint, we find for the non-mutating case that
	\begin{equation}\label{chronic_sstate}
	P^*(\vec{x}_0) \approx  \kappa\delta / \sigma; ~~~ L^*(\vec{y}_0) \approx \lambda / \beta.
	\end{equation}
	
	In non-mutating systems, this approximate steady state occurs with appropriate initial conditions, such as a large initial pathogen dose (solid green curve in Fig.~\ref{four-leaf}B). It is also quite similar to the chronic state reached by the mutating infection. Fig.~\ref{four-leaf}B shows the system trajectory for this mutating infection (solid black curve) through day 50, in comparison with two trajectories of the non-mutating infection (solid green curve and dashed black curve). This infection is a typical example of the valuable strategy of mutation allowing pathogens that would otherwise be cleared in small-dose initial conditions -- e.g. the example infection in Fig.~\ref{four-leaf} -- to avoid clearance (Eq.~\ref{cleared_sstate}) and converge to a chronic infection (Eq.~\ref{chronic_sstate}).

	The population density distribution of the pathogens during this process is shown in Fig.~\ref{evol50}A. The pathogens remain at their original location until the population nears its peak density, at which point several new strains arise farther from the control of the responding T-cells. It is these partial escape mutations which allow the resurgence of total pathogen density seen in Fig.~\ref{four-leaf}A just before day 10. Subsequent pathogen peaks in Fig.~\ref{four-leaf}A each have lower density, and Fig.~\ref{evol50}A shows that each of these successive peaks results in a decreased mutation distance in shape space. Because longer-distance mutations occur at lower rates, higher pathogen densities are required for them to occur.

  Since pathogen mutation can allow infections that would otherwise be cleared to become chronic if they can establish strains with low affinity to the initial immune response, we would expect greater success for pathogens with higher mutation rates, which are able to generate longer-distance mutations. Similarly, we would expect greater success for pathogens that face more specific immune responses, since they have a smaller range of T-cell recognition to evade. While these expectations are often borne out, our results also show deviations from this behavior, which are determined by both the underdamped oscillatory nature of the early-infection population dynamics and the strength of the resource constraint on the immune system.
  
\begin{figure}[ht!]
	\begin{center}
	\includegraphics[scale=1.0]{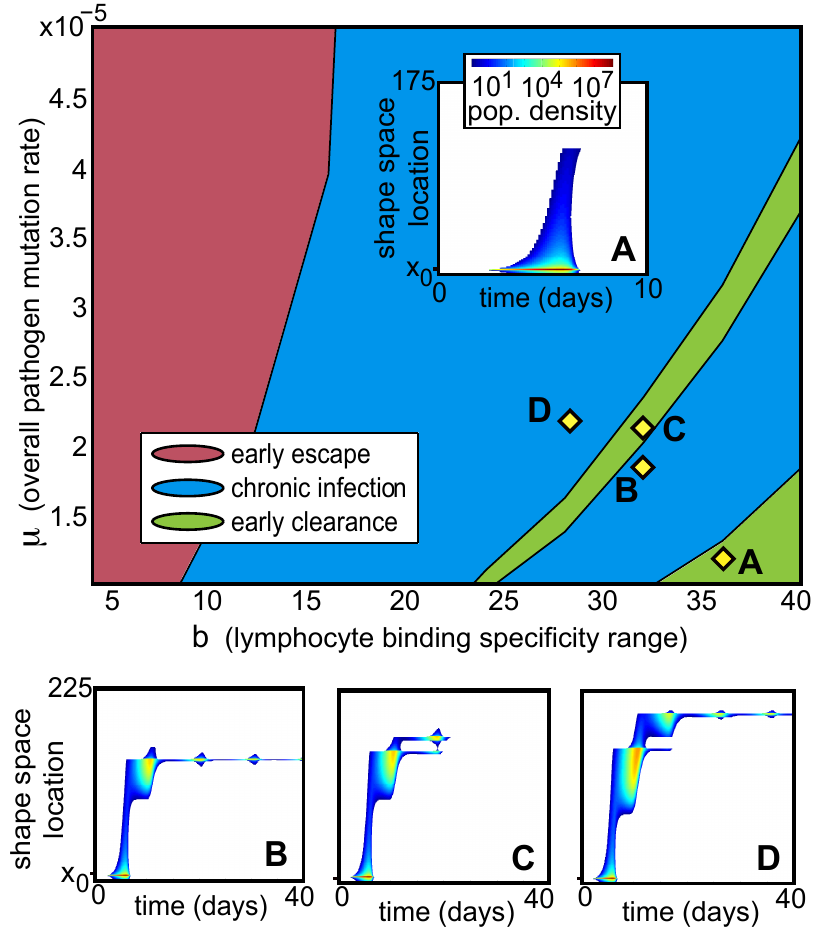}
	\end{center}
	\caption{ { \bf Phase diagram of deterministic infection trajectories under low homeostatic pressure, as a function of mutation rate $\mu$ and T-cell specificity range $b$. } In most cases, infections with a higher mutation rate in systems with narrower immune specificity range (i.e.~higher $\mu$ and lower $b$) are more successful at avoiding early clearance, but a band of cleared infections in the otherwise chronic regime highlights a trade-off: pathogen must mutate fast enough to evade an existing immune response, while keeping mutant numbers low enough to avoid stimulating another. A.~Shape space distribution of pathogen infection with a low mutation rate: the original inoculation and all mutants are killed by the initial immune response. B.~A faster mutation rate and narrower immune response allows pathogen mutants to avoid the initial immune response and establish a chronic infection. C.~As mutation rate increases and specificity range narrows, an increased number of pathogens avoid the initial immune response; this stimulates a more effective secondary response, which is able to clear the entire mutant population. D.~Chronic infection is established at a higher mutation rate when the secondary pathogen peak is able to establish mutants that can evade the secondary immune response. All infections have $\xi = 6.7\times10^{-5}~\mathrm{day}^{-1}$; other parameters are given in Table~\ref{params}. }
	\label{BCphasediagram}	
\end{figure}

\begin{figure}[ht!]
	\begin{center}
	\includegraphics[scale=1.0]{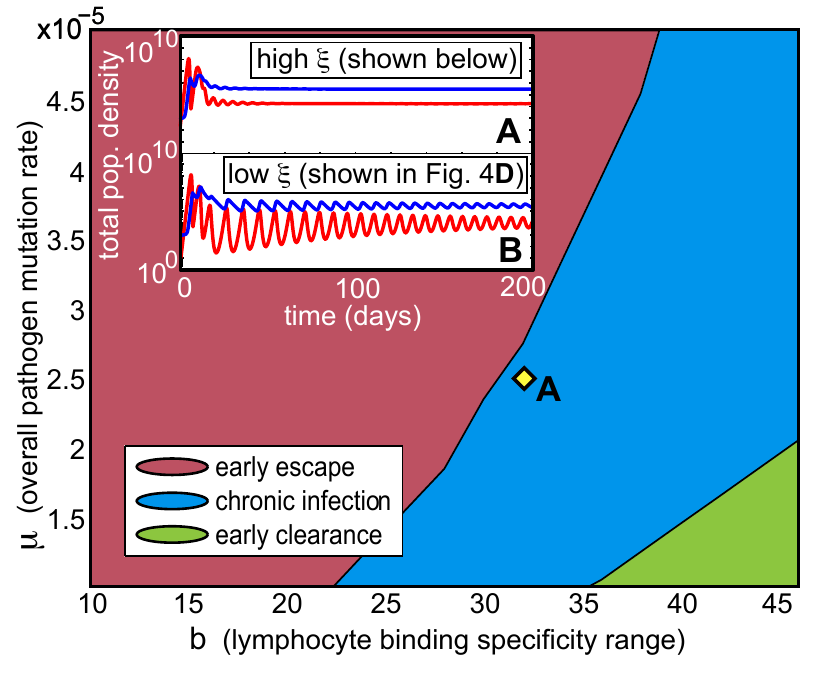}
	\end{center}
	\caption{ { \bf Phase diagram of deterministic infection trajectories under high homeostatic pressure. } Relative to Fig.~\ref{BCphasediagram}, this increased pressure ($\xi = 6.1\times 10^{-4}~\mathrm{day}^{-1}$ for all infections in the main phase diagram) has the overall effect of suppressing the immune response, allowing early escape and chronic infection to occur at lower pathogen mutation rates ($\mu$) and larger affinity ranges ($b$). In addition, as shown in the insets, convergence to the approximate steady state of chronic infection occurs much more quickly under high homeostatic pressure (inset A), due to the damping effect of the homeostatic constraint on population oscillations. Inset B shows population densities for an infection with low homeostatic pressure from Fig.~\ref{BCphasediagram}D; these display stronger, more sustained oscillations. Inset A and infection A have $\mu = 2.5\times 10^{-5}~(\mathrm{base}\cdot\mathrm{cycle})^{-1}$ and $b = 32$ sites. Inset B and its corresponding infection in Fig.~\ref{BCphasediagram}D have $\xi = 6.7\times 10^{-5}~\mathrm{day}^{-1}$, $\mu = 2.2\times 10^{-5}~(\mathrm{base}\cdot\mathrm{cycle})^{-1}$, and $b = 28$ sites. All other parameter values are specified in Table~\ref{params}. }
	\label{BCphasediagram2A}	
\end{figure}

  We examine the dynamics as a function of $\mu$, the pathogen mutation rate, and b, the width of the T-cells’ aﬃnity curve, at two different values of the homeostatic strength parameter $\xi$. Fig.~\ref{BCphasediagram} shows a phase diagram of system trajectories at a relatively low homeostatic pressure $\xi$. In most cases, our simple expectation is confirmed: infections with higher mutation rates and those that face more specific immune responses are those that become chronic rather than being cleared; and as $\mu$ grows and $b$ shrinks, the immune system becomes unable to control the pathogens at all. However, the details of the early-infection dynamics also give rise to a trade-off: to become chronic at a given immune system specificity, the infection needs a mutation rate both high enough to generate mutants out of the range of the initial immune response, and low enough to avoid stimulating too strong a secondary response. 
   
   This trade-off is apparent in the thin green band of clearance amid chronic infection outcomes in Fig.~\ref{BCphasediagram}. The pathogens are best able to become chronic when their mutation rate allows them to expand just far enough from the original strain to avoid being cleared (Fig.~\ref{BCphasediagram}B); the secondary immune response to these few surviving pathogens cells is too small to reduce them below the threshold, so the secondary T-cell and pathogen peaks converge to a chronic infection. At higher pathogen mutation rates, however, more mutants survive the original immune response; this can stimulate the secondary response strongly enough that it clears the secondary pathogen peak completely (Fig.~\ref{BCphasediagram}C), creating an isolated clearance band. At even higher mutation rates and narrower immune specificity ranges, the secondary pathogen peak is able to produce enough mutants of its own before being cleared to perpetuate the infection further (Fig.~\ref{BCphasediagram}D).
   
   The boundaries of these phase space regions depend in complex ways on the two evolving population distributions and on the dynamics of their interaction. In particular, boundaries between chronic infection and clearance, such as those creating the narrow clearance band in Fig.~\ref{BCphasediagram}, occur at phase space locations at which, at any point in time, the maximum single-site pathogen population density reaches the constant minimum pathogen threshold. If the maximum single-site density falls below the threshold at any time, the pathogen will be cleared everywhere; if it remains above the threshold at all times, the infection will survive. This threshold is most likely to be crossed near a minimum of total pathogen density; for example, reaching the threshold at the local pathogen density minimum following the primary (secondary) infection peak (see Fig.~\ref{four-leaf}A) corresponds to clearance by the primary (secondary) immune response, as shown in Fig.~\ref{BCphasediagram}B (D). At the low value of $\xi$ used in Fig.~\ref{BCphasediagram}, the infections in the green clearance band, including~\ref{BCphasediagram}C, escape the first response and succumb to the second. 
  
  At the relatively low homeostatic pressure strength used in Fig.~\ref{BCphasediagram} ($\xi = 6.7\times10^{-5}~\mathrm{day}^{-1}$), chronic infections' convergence to equilibrium typically takes several hundred days, with sustained oscillations in both populations (see Figs.~\ref{BCphasediagram}B,~\ref{BCphasediagram}D, and~\ref{BCphasediagram2A}B). The effects of increased homeostatic pressure ($\xi = 6.1\times10^{-4}~\mathrm{day}^{-1}$) are shown in Fig.~\ref{BCphasediagram2A}. Here the resource constraint, although still a small effect in comparison to other immune system interactions, has a significant damping effect on the oscillations in the early-infection population dynamics, reducing the time for infections to converge the approximate chronic steady state (Fig.~\ref{BCphasediagram2A}A). 
  These smaller oscillations keep the minimum population densities higher, impairing the immune system's ability to clear the pathogen population below its threshold. (Note that in reality stochastic effects, rather than a deterministic threshold, would determine the fate of a very small population; the values of $\mu$ and $b$ would merely determine the \textit{probability} of pathogen extinction.) Increased homeostatic pressure thus lowers the effectiveness of the overall immune response, allowing early escape and chronic infection to occur at lower pathogen mutation rates ($\mu$) and larger affinity ranges ($b$). In the crucial early stages of the infection that decide whether a rapid mutator will be fully cleared or converge to a chronic infection which may later do much more damage, the level of specificity in T-cell binding and the effect of the immune system's resource constraint can make a significant difference.
  
  Once chronic infection is reached, increased homeostatic pressure accelerates the narrowing of T-cell distribution by stimulating faster decay for all T-cells; this allows pathogen branching and eventual immune escape to occur much more quickly, in hundreds rather than thousands of days. Otherwise, the mechanisms leading to these chronic infection behaviors are similar at different values of $\xi$. For computational efficiency and easier visualization, the infections shown in Figs.~\ref{evol50},~\ref{4types}, and~\ref{four-leaf} thus use this higher homeostatic pressure; however, the discussion of branching and escape behaviors applies to infections over a range of these values.

\subsection*{Chronic Infection Dynamics}

  The chronic infection state that arises in a large class of infections displays particularly interesting dynamics, including instances of antigenic drift, evolutionary branching corresponding to asexual speciation of the pathogen population, and the eventual dramatic escape of the pathogens from immune system control.

  Throughout the chronic phase, both the mutation and high-dose no-mutation simulations show control of the pathogen population (without clearance) by the adaptive immune system. It has been suggested that chronic infections are more often controlled by innate immunity or resource constraints (both are well modeled by the carrying capacity term in Eq.~(\ref{dP-mut})), and that T-lymphocytes take on an unresponsive ``exhausted'' phenotype~\cite{stromb12_exhst}. However, it has recently been shown that HIV is an example of a chronic infection that is controlled by the adaptive immune system~\cite{deboer}, and other infectious diseases with immune evasion strategies may have similar dynamics.

	\subsubsection*{Antigenic drift}
	
  Initially, peaks in pathogen density drive bursts of mutation, but when the chronic phase is reached the evolution rate declines and stabilizes, as pathogen density converges to a near-constant level $P_{tot}\approx\delta \kappa/\sigma$. As long as the pathogen population at its peak remains large enough, however, some mutants are still spawned very near in shape space to the pathogen peak, and those further from the existing immune response have a competitive advantage. This results in a gradual traveling wave of pathogens in the phenotypic space, shown in Fig.~\ref{evol50}C. This type of motion through a one-dimensional fitness space has been seen in analytical studies of related models, including the specific situation of antigenic drift in RNA viruses~\cite{tsimring96_rnawaves}, and a pathogen's interaction with a traveling adaptive immune response~\cite{sasaki94_drift}. 
  
    The corresponding T-cell dynamics in Fig.~\ref{evol50} are purely the result of selection (we have ignored mutation of lymphocyte receptors, a choice more consistent with T-cell dynamics~\cite{janeway05_immunobiology}). Thus, wave-like motion driven by the interplay between mutation and selection does not appear. Instead, a constantly maintained low density of unactivated cells across shape space allows T-cells to peak wherever they are most highly stimulated, while homeostatic pressure keeps these peaks narrow by causing formerly stimulated T-cells to decay. The decay of formerly stimulated T-cells during a chronic infection scenario is consistent with findings that formation of long-lived antigen-independent memory T-cells is impaired during chronic infection, and a strong active memory is only formed following acute infection~\cite{wherry2004antigen}.

	\subsubsection*{Evolutionary branching}
	
  During this chronic phase, we also observe evolutionary branching corresponding to asexual speciation of the pathogen population, which is fundamentally driven by the coevolutionary interaction between the pathogens and the predatory immune response. This branching is visible in Fig.~\ref{4types}B and shown in more detail in Fig.~\ref{evol50}B.
  
  As the immune response controls the pathogens, the decay caused by operation above homeostatic equilibrium causes its density distribution in shape space to narrow, allowing pathogen strains to proliferate on the opposite side of the T-cell peak. This peak has higher affinity to the intermediate strains, and eventually clears them, leaving two separated groups of pathogens which evolve away from the immune response in opposite directions. 

  This speciation emerges as a direct result of the pathogen-immune interaction, without the typical drivers of allopatry (i.e.~spatial separation) or minima in the pathogen's fitness landscape. The model's assumption that all pathogens and T-cells are well-mixed means that pathogen speciation is sympatric, rather than the result of a varying spatial distribution within the population~\cite{doebeli00_branching,birkybarr09_asexspec,geritz04_ADofspec}. The underlying mechanism is the pathogen's ability to survive in two distinct niches, which arise due to the disruptive selection of the immune response on the phenotype continuum. Our results show a form of speciation that results directly from an interaction between predator and prey, especially an interaction mediated by a varying phenotypic trait.

  For several hundred days after the pathogen split, the joint stimulation of the two strains maintains a strong T-cell peak between them; however, eventually they separate enough for selection to favor two distinct T-cell peaks over a single central one (Fig.~\ref{evol50}C). The immune system's enhanced control of the pathogens here comes at the cost of approximately doubling the total T-cell density, causing T-cells throughout shape space to experience increased homeostatic pressure. 

	\subsubsection*{Eventual immune escape}
	
  At each branching event, both total populations increase, peaking slightly as the T-cell population splits. (This causes the cusps in total pathogen population in the inset in Fig.~\ref{four-leaf}A.) The homeostatic pressure of maintaining such a high-density immune response both impairs the ability of activated T-cells to maintain control of existing pathogen strains, and causes increased susceptibility in the unactivated T-cell regions. After almost 900 days of infection, this causes a dramatic immune escape (Fig.~\ref{evol50}) in which the pathogens grow out of control, spreading through the phenotypic space and proliferating to carrying capacity. Thus, the pathogen speciation contributes to the weakening and eventual breakdown of the immune system. This sudden increase in viral load after a long chronic period is reminiscent of the long-term qualitative features of infections such as HIV and SIV~\cite{janeway05_immunobiology,nowak_evolutionary_2006,mudd12_HIVvaccine} (Fig.~\ref{four-leaf}A inset). 
  
\section*{Discussion}  
  
  Overall, our model provides a method of investigating patterns and behaviors in the coevolutionary dynamics of the immune system, by abstracting the biochemical detail of the T-cell-pathogen interaction and representing it on a low-dimensional phenotypic shape space. With simple choices for the relations between binding shape, fitness, and relative mutation rates, the results reproduce the overall qualitative behavior of well-known chronic infections such as HIV and SIV. 
  
  Although the true shape space of binding phenotypes is likely not one-dimensional~\cite{smith97_ssdimens}, sequences of preferred or coordinated mutations~\cite{dahirel11_HIVlinkage} may effectively reduce its dimension, making our choice a reasonable approximation for many infections. However, a one-dimensional space does render independently evolving pathogen strains more likely to encounter each other again through antigenic drift. When this occurs, the overlap of stimulation regions incites a strong T-cell response between strains, as seen in Fig.~\ref{evol50} around day 700. This ensures that convergent evolution does not occur even in a one-dimensional shape space; the independent strains persist unless cleared by the immune response. However, interesting questions remain about the specifics of the model's behavior in higher-dimensional shape spaces, as well as the effect of making disease-specific assumptions about the fitnesses and mutation rates of particular pathogen strains.
  
   The question of the T-cell binding specificity needed to mount the most effective response has been the subject of much discussion and study\cite{ostperel79_ss,kovsmrlj10_HLAspecificityHIV,imm4phys}. Our results display a trade-off in na\"{i}ve cell specificity (Fig.~\ref{BCphasediagram}) that arises from a resource constraint, suggesting that the immune system's strategy for managing its limited resources could play an important role in its ability to control to a rapidly mutating infection. Another often-discussed reason for specificity trade-offs in na\"{i}ve T-cells is the necessity of avoiding self-reactivity~\cite{imm4phys}, an effect not included in this model.
  
   Evolutionary branching in similar predator-prey systems has been studied using the framework of adaptive dynamics (e.g.~\cite{geritz97_ADprl,doebeli00_branching,geritz04_ADofspec}). Like the shape space technique, these models represent populations on a phenotypic space, avoiding the complexity of high-dimensional models that track evolution at a genetic level. In these models, evolutionary branching occurs at a fitness minimum which is usually built into the model or dependent on interactions of the branching population with its environment. The branching in our model occurs at an effective pathogen fitness minimum created solely by the presence of predatory immune response, since the fitness landscape independent of the immune system is flat. The branching we see is thus driven by the coevolutionary interaction, and the exact point of branching depends on these dynamics.
  
  The increased immune system fragility observed during a chronic infection under homeostatic pressure, in which the immune system is highly specialized to control the existing pathogen strains at the cost of much lower protection in the unactivated regions, is typical of many complex systems displaying Highly Optimized Tolerance~\cite{stromb06_imsc,carlson02_pnascomplexity,zhou02_pnasHOTbio,zhou05_HOTevolution}. It leads to a state in which infection in new areas can proliferate with little control by adaptive immunity. This causes the suddenly increased viral load similar to that associated with the onset of AIDS~\cite{janeway05_immunobiology,nowak_evolutionary_2006}, and the failure of immune system control.

\section*{Acknowledgments}
This material is based upon work supported by the David and Lucile Packard Foundation, 
the Office of Naval Research MURI grants N000140810747 and 0001408WR20242, 
the Institute for Collaborative Biotechnologies through contract no.~W911NF-09-D-0001 from the U.S. Army Research Office, and the National Science Foundation Graduate Research Fellowship Program under Grant No.~DGE-1144085.

\bibliography{References_1}

\newpage


\section*{Appendix S1. Estimation and discussion of parameter values.}

This Appendix provides biological details and discussion on parameter choices used in this model. To obtain insight into the co-evolutionary dynamics of rapidly mutating pathogens and the adaptive immune response, we choose biologically relevant values for these parameters (given in Table 1 in the main text), drawing from previously measured and modeled characteristics of T-cells and well-studied viruses.
		
	The parameter values shown in Table~\ref{seanvalues} are adopted or slightly modified from a previously published shape space model of T-cell dynamics~\cite{stromb13_physbio} or a related model of thymic T-cell influx during chronic infection~\cite{stromberg12antia}, in order to reproduce the basic acute infection immune system dynamics seen in these models. The rate of pathogen growth $\lambda$ is consistent with~\cite{stromb13_physbio}, and similar to previously measured values of the initial expansion rate of HIV-1~\cite{little1999viral,ribeiro2010estimation}. The rate of pathogen killing by the immune response has been estimated to be between 0.1 and 0.5 $\mathrm{day}^{-1}$ near the peak of viremia in HIV and SIV infections~\cite{ganusov2006estimating,goonetilleke2009first}. The concentration of virus-specific T-cells responding to the primary infection near this peak can be roughly approximated as $10^4~\mu\mathrm{l}^{-1}$, based on measurements of the CD8 T-cell response to an infection of lymphocytic choriomeningitis virus (LCMV) in the mouse spleen~\cite{murali1998counting,de2001recruitment}). Given these estimates, a value on the order of $10^{-5}~\mu\mathrm{l}\cdot\mathrm{day}^{-1}$, as used in~\cite{stromb13_physbio} and~\cite{stromberg12antia}, is reasonable for a pathogen killing rate per day, per concentration unit of T-cells in the adaptive response.
	
	The T-cell replication and decay rates used here (denoted $\sigma$ and $\delta$, respectively) are consistent with those estimated from measurements of the immune response to LCMV in mice (replication rates between 1 and 3 $\mathrm{day}^{-1}$, decay rates between 0.3 and 0.5 $\mathrm{day}^{-1}$)~\cite{de2001recruitment,de2003different}. An approximate stimulation coeﬃcient $\kappa$ is given by early-stage HIV data in~\cite{davenport2004kinetics}, in which the significant T-cell response is observed to begin when viremia is on the order of $10^{5}~\mu\mathrm{l}^{-1}$.
	
	From a previous shape space modeling paper~\cite{stromb13_physbio}, we adopt an initial na\"{i}ve T-cell concentration $L_0$ of $3~\mu\mathrm{l}^{-1}$ for each shape space site (T-cell clone). This is a biologically reasonable value with typical overall na\"{i}ve T-cell concentrations. We can then estimate $\Gamma$, the na\"{i}ve cell inﬂux at a single shape space site (T-cell clone), from the initial equilibrium condition in the absence of pathogen, given in Eq. 10 in the main text: setting $L0 = \Gamma / \delta$ gives an approximate value of $\Gamma = 1~(\mu\mathrm{l}\cdot\mathrm{site}\cdot\mathrm{day})^{-1}$. We chose a value of $\phi = 10^{10}~\mu\mathrm{l}^{-1}$ for the pathogen carrying capacity in order to give acute infection dynamics comparable to those in~\cite{davenport2004kinetics} and~\cite{stromb13_physbio}.
	
\begin{table}[!ht]
	\renewcommand{\thetable}{S1}
\caption{ Parameters adopted from previously published models~\cite{stromb13_physbio,stromberg12antia}, and consistent with several measurement and estimation studies on immune system and virus properties (see text). }
	\begin{center}
	\begin{tabular}{| l | c | c |}
	\hline
	Parameter & Sym. & Value\\
	\hline
	pathogen growth & $\lambda$ & $3~\mathrm{day}^{-1}$ \\
	pathogen killing & $\beta$  & $10^{-5}~\mu\mathrm{l}\cdot\mathrm{day}^{-1}$ \\
	na\"{i}ve cell influx & $\Gamma$ & $1~(\mu\mathrm{l}\cdot\mathrm{site}\cdot\mathrm{day})^{-1}$ \\
	T-cell replication & $\sigma$ & $3~\mathrm{day}^{-1}$ \\
	T-cell decay & $\delta$ & $0.33~\mathrm{day}^{-1}$ \\
	stimulation coefficient & $\kappa$ & $10^{5}~\mu\mathrm{l}^{-1}$ \\
	\hline
\end{tabular}
\end{center}
\label{seanvalues}
\end{table}
	
	We express mutation in terms of the overall mutation rate $\mu$ of the pathogens, in the commonly used units of substitutions per DNA base per replication cycle. Each epitope has on the order of 10 amino acids, for an estimate of about 30 base pairs; thus, under the assumption that all base pairs in the pathogen genome mutate independently at identical rates, we approximate the expected number of mutations per cycle for a given epitope $\vec{x}$ as
	\begin{equation}\label{muappprox}
	\mu ' (\vec{x}) \approx 30\mu.
	\end{equation} 
	Consequently, if every member of a population of $P$ identical epitopes replicates once, we will have $P\times 30\mu$ mutants. In terms of our model, then, $\mu'(\vec{x})$ can be interpreted as the fraction of pathogens at location $\vec{x}$ that differentiate to locations $\vec{x}' \neq \vec{x}$, determined by the normalized mutation kernel $Q$ and the simulation space size $n$:
	\begin{equation}
	\mu ' (\vec{x}) = \sum_{\vec{x}'\neq\vec{x}} Q(\vec{x},\vec{x}') = \chi \left(\frac{2}{\pi}\right)^{\frac{1}{2}} \sum_{\vec{x}'\neq\vec{x}} |\vec{x}-\vec{x}'|^{-2}.
	\end{equation}
	Here, the sum is performed over all $n-1$ sites $\vec{x}'$ that are not equal to $\vec{x}$. For our chosen kernel $Q$, as given in Eq.~7 in the main text of the paper, and $n=400$, $\mu'$ is nearly constant across the simulation space (i.e., $<\mu'(\vec{x})>\approx \mu'(\vec{x}),~ \forall ~\vec{x}$), so we set its average value equal to our approximation in Eq.~\ref{muappprox}:
	\begin{equation}
	<\mu'(\vec{x})> = \frac{\chi}{n} \left(\frac{2}{\pi}\right)^{\frac{1}{2}} \sum_{\vec{x}}\sum_{\vec{x}'\neq\vec{x}} |\vec{x}-\vec{x}'|^{-2} \approx 30\mu.
	\end{equation}
	Thus, having chosen $Q$ and $n$, the value of the parameter $\chi$ is uniquely determined from $\mu$. In this study, we use the standard value $\mu = 2.2\times10^{-5}$ substitutions per base pair per cycle, similar to commonly measured values for the rapidly mutating virus HIV-1~\cite{sanjuan2010viral}, and study the effects of variation between $1\times10^{-5}$ and $5\times10^{-5}$.
	The binding specificity parameter $b$ gives the width of the Gaussian affinity curve in shape space, setting the specificity of the na\"{i}ve T-cell response. We choose $b$ to correspond to typical values of the precursor frequency, or the probability of two different na\"{i}ve T-cells responding to the same epitope (about $1/10^5$~\cite{imm4phys,precursor2}). Starting from the uniformly distributed initial condition in which $L(\vec{y}) = \Gamma / \delta~\forall~\vec{y}$ (Eq.~10 in the main text of the paper), the number of na\"{i}ve T-cells under the affinity curve centered at some $\vec{y_0}$, and thus cross-reactive with the response to a pathogen maximally complementary to $\vec{y_0}$, is given in one dimension by
	\begin{equation}
	\frac{\Gamma}{\delta}\int\gamma(\vec{y},\vec{y_0})~d\vec{y} = \frac{\Gamma}{\delta}\int\exp\left[-\frac{|\vec{y}-\vec{y_0}|^2}{2b^2}\right]~d\vec{y} =  \frac{b\Gamma\sqrt{2\pi}}{\delta}.
	\end{equation}
	Although we only simulate a small portion of the full shape space for computational efficiency, the total number of T-cells in our system is approximated by the initial na\"{i}ve density times the total number of clones in the human immune system, with reasonable values between $10^{6} - 10^{7}$~\cite{imm4phys}. To estimate $b$, we can then set the fraction of cross-reactive na\"{i}ve T-cells equal to the precursor frequency:
	\begin{equation}
	b\sqrt{2\pi}/(5\times10^{6}) = 10^{-5} ~~~\longrightarrow~~~ b \approx 20.
	\end{equation}
	In this paper, we investigate values of $b$	corresponding to the above range of total na\"{i}ve T-cell clones in the system.
	
   We investigated the system under several values of $\xi$, which controls the strength of the damping homeostatic pressure. We chose values small enough that $\xi << \sigma R /|L_{tot}-R|$ and $\xi << \delta R /|L_{tot}-R|$, so that the homeostatic pressure would not significantly alter the expected chronic infection steady-state levels. In the main text of the paper we present results obtained at two different values of $\xi$. The higher value (Fig. 5 in main text) shows much stronger damping of total population density oscillations in acute infection, and converges to approximate chronic equilibrium quickly. Infections at the lower value of $\xi$ (Fig. 4 in main text) take longer to converge to a chronic infection, branch, and eventually escape control. The branching and escape behaviors are observed consistently during chronic infections across values of $\xi$, and appear to be driven by the same mechanisms, although the timescale on which they occur changes, as well as the precise values of other parameters needed to generate a chronic infection. 
   
   In order to ensure that our choice to simulate only a limited region of the total shape space of na\"{i}ve T-cell clones does not substantially affect our results, we investigated the behavior of several infections in a shape space with a substantially larger T-cell population (between $10^{6} - 10^{7}$ na\"{i}ve clones). We found that extending the shape space in this way is effectively the same as lowering the parameter $\xi$: as discussed in the main text, the timescales of convergence to chronic infection, branching, and eventual immune escape are lengthened in a large shape space, while the overall effectiveness of the immune system is enhanced. However, the mechanisms leading to chronic infection, branching, and eventual pathogen escape are similar, though occurring on different timescales. In the main text and figures, we thus focus on presenting results from a small portion of shape space for computational efficiency and more effective visualization.
   
\end{document}